\documentclass[%
 aip,
 floatfix,
 amsmath,amssymb,
preprint,%
 reprint,%
]{revtex4-1}

\usepackage{graphicx}
\usepackage{dcolumn}
\usepackage{bm}

\usepackage[utf8]{inputenc}
\usepackage[T1]{fontenc}
\usepackage{mathptmx}
\usepackage{etoolbox}
\usepackage[per-mode=repeated-symbol]{siunitx}

\renewcommand{\selectlanguage}[1]{}
\makeatletter
\def\@email#1#2{%
 \endgroup
 \patchcmd{\titleblock@produce}
  {\frontmatter@RRAPformat}
  {\frontmatter@RRAPformat{\produce@RRAP{*#1\href{mailto:#2}{#2}}}\frontmatter@RRAPformat}
  {}{}
}%
\makeatother
\begin{document}

\preprint{AIP/123-QED}

\title[Highly versatile, two-color setup for high-order harmonic generation using spatial light modulators]{Highly versatile, two-color setup for high-order harmonic generation using spatial light modulators}
\author{A-K. Raab\textsuperscript{*}}
\email[]{ann-kathrin.raab@fysik.lth.se}
\affiliation{Department of Physics, Lund University, P.O. Box 118, 22100, Lund, Sweden}

\author{M. Schmoll}
\affiliation{Physikalisches Institut, Albert-Ludwigs-Universität Freiburg, Hermann-Herder-Straße 3, 79104, Freiburg, Germany}

\author{E. R. Simpson}
\affiliation{Department of Physics, Lund University, P.O. Box 118, 22100, Lund, Sweden}

\author{M. Redon}
\affiliation{Department of Physics, Lund University, P.O. Box 118, 22100, Lund, Sweden}

\author{Y. Fang}
\affiliation{Department of Physics, Lund University, P.O. Box 118, 22100, Lund, Sweden}

\author{C. Guo}
\affiliation{Department of Physics, Lund University, P.O. Box 118, 22100, Lund, Sweden}

\author{A-L. Viotti}
\affiliation{Department of Physics, Lund University, P.O. Box 118, 22100, Lund, Sweden}

\author{C. L. Arnold}
\affiliation{Department of Physics, Lund University, P.O. Box 118, 22100, Lund, Sweden}

\author{\\A. L'Huillier}
\affiliation{Department of Physics, Lund University, P.O. Box 118, 22100, Lund, Sweden}

\author{J. Mauritsson}
\affiliation{Department of Physics, Lund University, P.O. Box 118, 22100, Lund, Sweden}

\date{\today}

\begin{abstract}
We present a novel, interferometric, two-color, high-order harmonic generation setup, based on a turn-key Ytterbium-doped femtosecond laser source and its second harmonic. Each interferometer arm contains a spatial light modulator, with individual capabilities to manipulate the spatial beam profiles and to stabilize the relative delay between the fundamental and the second harmonic. Additionally, separate control of the relative power and focusing geometries of the two color beams is implemented to conveniently perform automatized scans of multiple parameters. A live diagnostics system gives continuous information during ongoing measurements.
\end{abstract}

\maketitle

\section{Introduction}
Sources of extreme ultra-violet (XUV) attosecond pulses, derived from high-order harmonic generation (HHG) in gases, hold considerable importance for basic research, including time-resolved spectroscopy with attosecond precision \cite{kraus_perspectives_2018}, and serve practical uses such as metrology tools in the semiconductor sector  \cite{den_boef_optical_2013, kinoshita_development_2014}.
However, due to its high nonlinearity, HHG is an inefficient process with typical conversion efficiencies on the order of $10^{-6}$ and lower \cite{weissenbilder_how_2022}.
To optimize the HHG flux, the microscopic interaction between the laser and a single atom \cite{corkum_plasma_1993, schafer_above_1993} as well as macroscopic propagation effects in the nonlinear interaction medium \cite{constant_optimizing_1999} have to be taken into account.
The single-atom response is mainly governed by laser parameters, such as peak intensity, wavelength and pulse duration.
It is also strongly affected by the use of multiple frequencies, such as the second or third harmonic in addition to the fundamental. 

The first two-color HHG experiments were performed already in 1993 \cite{perry_high-order_1993,watanabe_two-color_1994} following theoretical predictions \cite{schafer_phase-dependent_1992}.
Watanabe \textit{et al.}~obtained an increase of the harmonic yield by several orders of magnitude by adding the third harmonic of the driving field, thus opening an extremely promising route to optimize HHG \cite{watanabe_two-color_1994}. In fact, it was later predicted that the optimum electric field shape, which gives a high cutoff but also a high yield can be achieved by a multi-color synthesis \cite{chipperfield_ideal_2009}.
Multi-color HHG has been studied for many purposes ranging from fundamental understanding of the HHG process \cite{dudovich_measuring_2006, ishii_quantum_2008, dudovich_subcycle_2009, shafir_atomic_2009, fies_attosecond_2011} to  yield optimization \cite{kim_highly_2005, liu_significant_2006, ganeev_enhancement_2009, brugnera_enhancement_2010, roscam_abbing_divergence_2021}, orbital-angular momentum studies \cite{dorney_controlling_2019, gauthier_tunable_2017} and spectral \cite{mansten_spectral_2008, wei_selective_2013, wei_efficient_2014, mitra_suppression_2020} and temporal \cite{mauritsson_attosecond_2006} shaping. 

Experimentally, multi-color HHG is not easy to realize. Even in the simplest case with only two different colors in the driving field, one needs to obtain precise control over the relative field intensities, their relative delay with sub-cycle stability and their individual focusing geometries. 
An additional, recurring question for flux optimization is how to compare rigorously single-color and multi-color HHG.
 
In this work, we describe an experimental setup to generate high-order harmonics with a combination of the fundamental field and a fraction of its second harmonic. The setup is designed to obtain a precise parameter control, with the help of live diagnostics.
First, we  introduce the principles of multi-color HHG physics to identify the most important  parameters for the experiment. A small review of possible experimental realizations is presented where we explain the advantages and drawbacks of certain design choices to motivate the relevance of our setup. Second, we describe an experimental setup which uses a commercial laser with a pulse duration of $\SI{180}{\femto\second}$  at a central wavelength of $\SI{1030}{\nano\meter}$ together with its second harmonic.
The integration of  spatial light modulators (SLMs) gives us full control of the focusing conditions of the individual two color beams.
Furthermore, one SLM is used to actively stabilize the phase delay between the two colors with a stability of $<\SI{150}{\milli\radian}$. Finally, we present a selection of HHG results highlighting the versatility of our setup.

\section{Design of two-color high-order harmonic generation}

 
\begin{figure*}[t]
    \centering
    \includegraphics{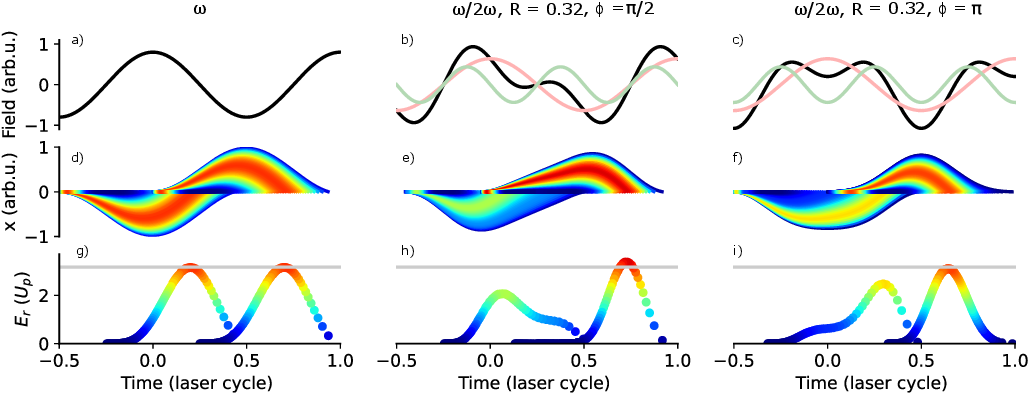}
    \caption{a)-c) Driving electric field, d)-f) returning electron trajectories and g)-i) return energies  for conventional one color  ($\omega$) and two-color HHG ($\omega/2\omega$). For the latter, an intensity ratio of 0.32 and a phase offset of $\pi/2$ (b), e) and h)) and $\pi$ (c), f) and i)) for the individual colors was chosen. The resulting electric field as an addition of the two colors (red, green) is drawn in black in the first row. The trajectories of the electrons that are ionized at a specific time in the laser cycle are color coded with respect to their return energy, displayed in the last row in units of the ponderomotive energy $U_p$. The single-color classical cutoff is highlighted by a grey line.}
    \label{fig:multicolor_schematics}
\end{figure*}

\subsection{Introduction to two-color high-harmonic generation} \label{ssec:HHGintro}
HHG is a laser-based source of coherent secondary radiation, discovered three decades ago. When intense ultrashort pulses are focused into a gas target, discrete spectral lines corresponding to odd high-order harmonics of the driving laser frequency $\omega$ are generated\cite{mcpherson_studies_1987, ferray_multiple-harmonic_1988}. This process can be understood physically by the three-step model \cite{corkum_plasma_1993, schafer_above_1993}.
In step one, due to a strong distortion of the atomic potential by the laser field, an electron can tunnel-ionize to the continuum.
It is then driven away by the laser field.
When the laser field changes sign, the electron is driven back to the parent atom (step two).
It may then recombine back to the ground state 
and its excess energy is emitted as a photon, typically in the XUV spectral range (step three).
This process is repeated every half-cycle of the driving field.
The spectrum of odd-order harmonics, typically associated with HHG, follows from the spectral interference of the mutually coherent emissions from every half-cycle.
In the presence of the second harmonic, the symmetry between consecutive half-cycles is broken, leading to the generation of also even-order harmonics.
The predictions of the three-step model for one and two-color HHG with the same linear  polarization are presented in Fig.~\ref{fig:multicolor_schematics}. 
The driving field is shown in a)-c) as a black solid line. In b) and c) the contributions of the fundamental (F, red) and second harmonic (SH, green) are indicated.

We introduce the  ratio $R$ between the second harmonic intensity $I_{\text{SH}}$ and total intensity $I_{\text{tot}}$:
\begin{equation}
    R = \frac{I_{\text{SH}}}{I_{\text{SH}}+I_{\text{F}}} = \frac{I_{\text{SH}}}{I_{\text{tot}}}\,.
\end{equation}
The resulting electric field $E$ depends on $R$ and on the relative two-color phase $\phi$ as
\begin{equation}
    E=\sqrt{(1-R)I_{\text{tot}}}\cos(\omega t) + \sqrt{RI_{\text{tot}}}\cos(2\omega t+\phi).
\end{equation}

The fields shown in Fig.~\ref{fig:multicolor_schematics} b) and c) are obtained for ${R=0.32}$ and for $\phi = \pi/2$  and $\phi = \pi$, respectively. 
The electrons that are born into the continuum at different ionization times follow different trajectories, which are illustrated in d), e) and f). 
We only show the trajectories of electrons that return to the parent atom and color-code their acquired kinetic energy. 
The kinetic energy upon recombination is shown in g)-i) as a function of the return time. 
 The maximum energy $E_{\text{max}}$ that can be achieved (dark red) depends on the ionization potential $I_p$ and is called the classical cut-off \cite{corkum_plasma_1993}.
 For a one-color field, 
\begin{equation}\label{eq:cutoff}
    E_\text{max} = I_p + 3.17U_p,
\end{equation}
where the ponderomotive potential $U_p$ is expressed as a function of the peak intensity $I$, the  central frequency $\omega$, the electron mass $m_e$ and charge $e$, the speed of light $c$, and the vacuum permittivity  $\epsilon_0$ as
\begin{equation}
    U_p = \frac{e^2}{2c\epsilon_0 m_e} \cdot \frac{I}{\omega^2}.
\end{equation}
In the last row of Fig.~\ref{fig:multicolor_schematics}, this single-color cutoff is highlighted by a gray line. 

The distribution of return energies is distorted by the addition of the second harmonic and strongly depends on the values for $R$ and $\phi$.
This can lead to a  cut-off energy higher than in the single-color case [Eq.~(\ref{eq:cutoff})] at the same total intensity [see Fig.~\ref{fig:multicolor_schematics} h)]
 as well as to the appearance of flat regions in the return energy versus time, where the yield is expected to be high.

\subsection{Possible experimental realizations}
Experimentally, two-color HHG can be very difficult  to realize as one needs to first perform waveform-synthesis and then obtain control over the relative phase, the relative field amplitudes and the individual focusing geometries. For this reason, most studies have focused on a narrow parameter space.
We will review and comment on the design choices of each of the above mentioned experimental challenges to motivate our own setup implementation. 

\subsubsection{Two-color field synthesis}
\begin{figure}
    \centering
    \includegraphics{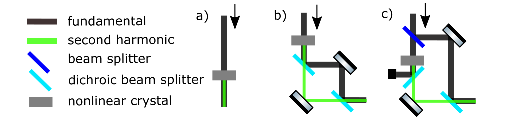}
    \caption{Schematics for different types of two-color waveform synthesizers: a) inline and b) and c) interferometric.}
    \label{fig:laser_schematics}
\end{figure}


In Fig.~\ref{fig:laser_schematics} different schemes for two-color field synthesis are shown. The easiest to implement and the most compact solution experimentally is the inline option (a)), where a nonlinear crystal is inserted into the beam path. This method is widely used \cite{perry_high-order_1993, schumacher_phase_1994, kim_highly_2005, dudovich_measuring_2006, oishi_generation_2006, doumy_attosecond_2009, ganeev_enhancement_2009, zheng_dynamic_2009, brugnera_enhancement_2010, brugnera_trajectory_2011, fies_attosecond_2011, raz_spectral_2012, shafir_resolving_2012, brizuela_efficient_2013, soifer_spatio-spectral_2013, wei_selective_2013, soifer_studying_2014, wei_efficient_2014, facciala_probe_2016, roscam_abbing_divergence_2020, gindl_attosecond_2023}.  
The desired intensity ratio between the fundamental and the harmonic can be obtained by choosing  the crystal thickness and phase-matching angle. Limitations might occur when using very short pulses, where the phase-matching bandwidth of the material will limit the spectrum and therefore pulse duration of the harmonic radiation. 
Another drawback is the impossibility to control the beam size, the field intensities and the  waveform shapes individually. 

This individual control can be achieved by an interferometric approach instead, as illustrated in Fig.~\ref{fig:laser_schematics} b) and c).
The most energy efficient one at first glance is obtained by placing the nonlinear crystal into the beam and then splitting the colors with dichroic optics, as depicted in b).
A slightly different version is shown in Fig.~\ref{fig:laser_schematics} c).
Here, the beam is split first with a fixed ratio, and the nonlinear crystal is placed in one of the arms.
In this setting, the crystal can be optimized to yield the best harmonic output in terms of efficiency and beam quality.
Also, impairment of the fundamental beam, e.g. depletion in the center, is avoided.
In both interferometer versions, each arm contains a single color and allows individual beam manipulation. 

When the second harmonic is generated in a nonlinear crystal co-linearly, as shown in all examples in Fig.~\ref{fig:laser_schematics}, it will be perpendicularly polarized relative to the fundamental. If parallel relative polarization is desired, a half-wave ($\lambda/2$) plate designed for one color can be added to the beam path. 

\subsubsection{Focusing geometry}

The focus waist $w_0$ of a Gaussian beam when focused with a focal length $f$, a beam radius $W$, wavelength $\lambda$ and a beam quality parameter $M^2$ is
\begin{equation}\label{equ:w0}
    w_0 = \frac{M^2\lambda f}{\pi W} \, .
\end{equation}
The Rayleigh length has the same wavelength scaling factor:   
\begin{equation}\label{equ:zr}
    z_R = \frac{\pi w_0^2}{M^2 \lambda} = \frac{M^2\lambda f^2}{\pi W^2} \,
\end{equation}
The beam size of the second harmonic will be a factor of $\sqrt{2}$ smaller, if second harmonic generation results from a collimated fundamental beam. In an inline setup this would then result in a focus size difference of ${\sqrt{2}/2}$ [Eq.~(\ref{equ:w0})] but equal Rayleigh length [Eq.~(\ref{equ:zr})] compared to the fundamental. An interferometric setup allows independent beam size management to achieve the desired values for $z_R$ and $w_0$. 
\subsubsection{Relative phase control}
To control the relative phase between the two fields in an inline setup, one has to rely on changing the amount of material the light goes through. Most works use a thin plate of fused silica at Brewster angle where small deviations from this angle will delay the components differently. While many publications do not comment at all on the introduced spatial beam displacement by transmitting through an angled plate, Brugnera \textit{et al.} concluded that as long as the displacement is significantly smaller than the medium width and length, it will not have an influence on the experiment \cite{brugnera_trajectory_2011}. Eliminating the displacement issue can be done using a wedge-pair with variable insertion instead \cite{doumy_attosecond_2009,soifer_studying_2014, zhai_ellipticity_2020,  gindl_attosecond_2023}. 
In an interferometric setup the relative phase control can be easily realized by adding a delay stage into one arm which is then not affecting the spatial properties \cite{calegari_efficient_2009, wirth_synthesized_2011}.

\subsubsection{Relative intensity control}
Inline-setups are implemented most straight-forwardly, but provide very limited options to carefully control the relative intensities, except for turning the nonlinear crystal to vary the conversion efficiency.  
In an interferometer, the options for attenuating pulse energies to change $R$ or the total intensity must be carefully considered: when the pulses are sufficiently long ($> \SI{30}{\femto\second}$) an attenuator consisting of a half wave plate and a polarizer can work very well. For shorter pulses, when high reflectivity thin film polarizers do not support a sufficiently large  bandwidth, an iris is often used. This, however, affects not only the transmitted energy but also the beam size before the final focusing optics, and consequently 
the focus waist sizes. In experiments where exploring different ratios of the two colors is not the main goal, aperturing one of the beams can be used to match the focus sizes of the individual colors \cite{ishii_quantum_2008}. 

\subsubsection{Design choices for a setup prioritizing flexibility and parameter control}

In summary, two-color setup design and diagnostic capabilities highly depend on the purpose and on which complications mentioned above can be neglected.
In our work, we prioritize flexibility and parameter control, thus choosing a setup with a configuration as shown in Fig.~\ref{fig:laser_schematics} c).
In the next section, we will present our interferometric setup with the possibility to shape the focusing geometry, intensity and phase relation between both colors by integrating SLMs.

\section{Flexible interferometric $\omega/2\omega$ high-order harmonic generation}

\subsection{Experimental setup}\label{subsec:setup}
\begin{figure}
    \centering
    \includegraphics{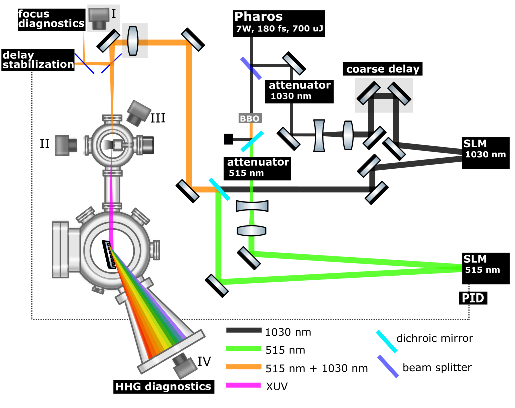}
    \caption{Sketch of the experimental setup. The light gray boxes represent motorized translation stages.}
    \label{fig:setup_new}
\end{figure}
Our experimental setup is depicted in Fig.~\ref{fig:setup_new}.
The output of a compact, turn-key Ytterbium-doped laser system with a pulse duration of $\SI{180}{\femto\second}$  and pulse energy of $\SI{700}{\micro\joule}$ at a repetition rate of $\SI{10}{\kilo\hertz}$ (Pharos, Light Conversion) is divided by a beam splitter.
$\SI{80}{\percent}$ of the power is reflected into the $\SI{1030}{\nano\meter}$ (black) interferometer arm, $\SI{20}{\percent}$ is transmitted.
A $\SI{1}{\milli\meter}$ thick BBO crystal converts the transmitted light to $\SI{515}{\nano\meter}$ (green). A dichroic mirror reflects the residual infrared (IR) into a beam dump. 
In both interferometer arms, a combination of a motorized $\lambda/2$ plate and a thin-film polarizer acts as an attenuator to continuously tune the transmitted power. After the attenuators both colors are horizontally polarized. To use the whole area of the SLMs ($\SI{9.6}{\milli\meter} \times \SI{15.36}{\milli\meter}$, Santec SLM300), and to decrease the intensity to avoid damage, the beams are enlarged via lens telescopes from a $1/e^2$ diameter of $\SI{4}{\milli\meter}$ to $\SI{8}{\milli\meter}$. Both colors are recombined through another dichroic mirror and are focused with an $f=\SI{+250}{\milli\meter}$ lens into a gas jet (backing pressure of $\SIrange{2}{4}{\bar}$ and gas nozzle diameter of $\SI{42}{\micro\meter}$), which can be moved in three directions.
The generated XUV radiation is sent via an imaging flat-field grating onto a multichannel plate (MCP) where the phosphor screen behind is recorded with a camera, denoted by IV in Fig.~\ref{fig:setup_new}.
A very thin glass plate close to  Brewster angle before the vacuum chamber reflects approximately $\SI{1}{\percent}$ of the power into the delay stabilization and focus diagnostic parts.   
Just in front of the vacuum chamber, the maximum power for $\SI{1030} {\nano\meter}$ is $\SI{4.5}{\watt}$ corresponding to a pulse energy of $\SI{450}{\micro\joule}$ and for $\SI{515}{\nano\meter}$  $\SI{0.35}{\watt}$ corresponding to $\SI{35}{\micro\joule}$. The split of the maximum pulse energy values can be varied by simply replacing the beam splitter at the beginning of the interferometer.

\subsection{Focus control employing SLMs}
\begin{figure}
    \centering
\includegraphics{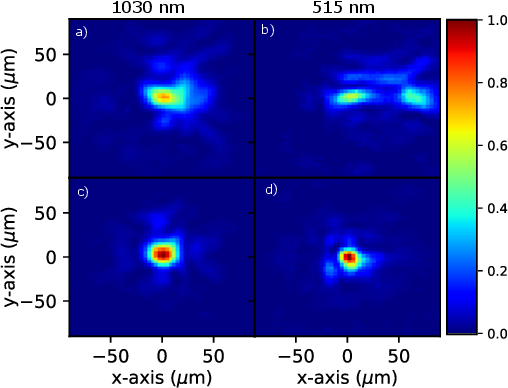}
    \caption{Uncorrected foci of a) the infrared  and b) second harmonic compared to c) and d)  after a correction is applied on the respective SLMs.}
    \label{fig:SLMs}
\end{figure}
The two-color beam is focused into the HHG chamber using a plano-convex lens, which introduces  chromatic aberations resulting in different focus positions for the two colors along the propagation direction. Additionally, a non-perfect re-collimation after the enlargement lens telescopes in the two interferometer arms influences focus position and size, as well as aberrations introduced by the large number of transmissive optics. 

For diagnostics, a part of the beam is reflected out by the thin plate directly in front of the vacuum chamber and focused onto the chip of a camera, denoted by I in Fig.~\ref{fig:setup_new}.
This camera is mounted on a motorized translation stage, allowing to take images along the propagation direction. 
Images captured a few millimeters before and after the focal plane can be used to obtain the beam quality factor $M^2$ as well as to retrieve the wavefront, using a modified version of the Gerchberg-Saxton algorithm \cite{gerchberg_practical_1972, allen_phase_2001}.
With the help of numerical beam propagation, the wavefront can be obtained in any plane of interest and aberrations can be corrected by displaying the inverse of the retrieved wavefront in the plane of the SLM.
In Fig.~\ref{fig:SLMs}, our generation foci are shown without active wavefront shaping for $\SI{1030}{\nano\meter}$ a) and $\SI{515}{\nano\meter}$ b).
The results of this correction for both colors are shown in Fig.~\ref{fig:SLMs} c) and d).
The images before and after correction share the same colorscale, highlighting the significant improvement of the peak intensity in the focus.

We measured focus sizes equal to $w_{0,\omega} = \SI{27}{\micro\meter}$ and $w_{0,2\omega} = \SI{14}{\micro\meter}$, in excellent agreement with the predicted beam sizes, see Eq.~(\ref{equ:w0}).
Adding a tilt to the wavefront in the $x$ and $y$ directions, perpendicular to the propagation axis,  allowed for small alignment corrections. To move the beams independently of each other along the propagation direction, a parabolic phase front is applied,  resulting in minimal beam focusing or defocusing after the SLM without influencing $w_0$ significantly. Scanning camera I in Fig.~\ref{fig:setup_new} on the translation stage allows the  determination of both foci positions relative to each other as well as their size and the overall beam quality at all times. 

\begin{figure}
    \centering
    \includegraphics{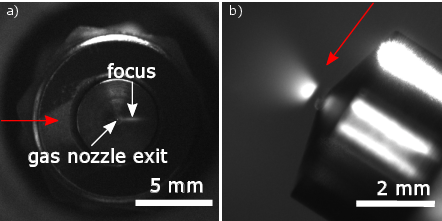}
    \caption{a) Side-view of the gas nozzle and  b)  top view of the gas nozzle.The laser direction is illustrated by a red arrow. Those views correspond to II and III in Fig.~\ref{fig:setup_new}.}
    \label{fig:cameras_nozzle}
\end{figure}

To estimate the focus position relative to the gas target inside the vacuum chamber, two cameras are used to image the proximity of the gas target through chamber windows from the side and from the top, denoted in Fig.~\ref{fig:setup_new} by II and III, respectively. In Fig.~\ref{fig:cameras_nozzle}, the side a) and top views b) are shown. 
These views serve the purpose of visualizing the alignment of the gas nozzle relative to the laser focus. 
In Fig.~\ref{fig:cameras_nozzle} a) an example is shown where the pressure in the vacuum chamber is  $\approx \SI{0.1}{\milli\bar}$, which is sufficient to ionize the remaining gas. Using this method, one can calibrate with high precision the position of the focus waist relative to the gas target. The longitudinal focus position can then be changed either by moving the focusing lens or by applying a parabolic phase on the SLMs. 

Camera III in Fig.~\ref{fig:setup_new} shows the gas target from above, in this example, under ambient pressure around ${\SI{1e-6}{\milli\bar}}$. 
From the image in Fig.~\ref{fig:cameras_nozzle} b), one can measure the exact distance between the nozzle exit and the laser focus. 

\subsection{Phase control}
\begin{figure}
    \centering
    \includegraphics{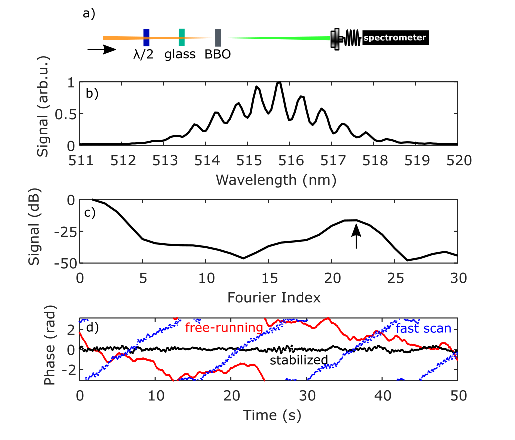}
    \caption{a) The setup of the delay stabilization. b) Spectrum of the two overlapped green components, showing the interference fringes. c) Fourier Transform of b). d) Comparison of the relative two-color phase: free-running (red), actively stabilized long term (black) and  actively stabilized for fast phase scans (blue).}
    \label{fig:diagnostics_phase}
\end{figure}

The stabilization of the delay between the $\SI{1030}{\nano\meter}$ and the $\SI{515}{\nano\meter}$ arm uses the beam transmitted through the second thin plate in the diagnostic part of the setup.
The box marked as "Delay stabilization" in Fig.~\ref{fig:setup_new} is presented explicitly in Fig.~\ref{fig:diagnostics_phase} a).
Both colors first go through a $\lambda/2$ plate designed for $\SI{1030}{\nano\meter}$, which rotates the polarization of the $\SI{1030}{\nano\meter}$ radiation by $\SI{90}{\degree}$, but leaves the $\SI{515}{\nano\meter}$ radiation  practically untouched.
Transmission through $\SI{9}{\milli\meter}$ of BK-7 glass introduces a delay between the two-color components.
The second harmonic of the fundamental pulse is generated in a BBO.
Both second harmonic components have the same polarization and spectrally interfere. The spectral  fringes are  recorded by a commercial fiber spectrometer and are shown in Fig.~\ref{fig:diagnostics_phase} b). The recorded spectrum is Fourier transformed c), and the phase is extracted at the position of the peak marked by an arrow. This phase is plotted in Fig.~\ref{fig:diagnostics_phase} d) over a span of $\SI{50}{\second}$. It is used as a feedback to the $\SI{515}{\nano\meter}$ SLM. A  proportional–integral–derivative (PID) control  calculates the  phase correction which is necessary for active stabilization. The SLM allows a phase variation of $2\pi$, with a response time of  $\approx\SI{200}{\milli\second}$. 

In Fig.~\ref{fig:diagnostics_phase} d) the free-running, unstabilized case (red) is compared to the actively stabilized case (black) over the same time span. The free-running phase-drift is slow compared to the stabilization response time. The standard deviation of the stabilized phase is $<\SI{150}{\milli\radian}$. During each measurement,  the phase is logged to allow finer data sorting later for even higher accuracy. This is especially useful if the setup is operated in conditions where the phase is changed very rapidly.
Such a case is presented in Fig.~\ref{fig:diagnostics_phase} d) in blue, where the intensity ratio of fundamental to green was changed from below $\SI{1}{\percent}$ to $\SI{30}{\percent}$ for different fast phase scans consecutively. Depending on the experimental conditions, the fringe contrast can be optimized by placing selective filters for IR or green in front of the spectrometer shown in Fig.~\ref{fig:diagnostics_phase} a). Alternatively the BBO crystal can be moved along the beam focus to produce more or less second harmonic. There were no issues in phase stability as long as the Fourier peak signal was above $\approx \SI{-25}{\decibel}$.
The phase measured this way is purely relative. To obtain an absolute phase reference for the  experiment, one can calibrate it by using, for example, the minimum in harmonic yield \cite{roscam_abbing_divergence_2021}. 

\subsection{$\omega/2\omega$ ratio control}
To control the total field intensity in the gas target as well as the ratio between the two colors, each interferometer arm contains an attenuator with a half-wave plate on a motorised rotation mount and a $\SI{45}{\degree}$ thin film polarizer (TFP). 
The beam size of the fundamental and second harmonic are measured using camera I in Fig.~\ref{fig:setup_new} for each experimental configuration and an $M^2$ fit is performed. With this knowledge, automated scans are possible at a constant total peak intensity ${I_{\text{tot}}=I_{\text{SH}}+I_{\text{IR}}}$ while changing the contribution of the individual fields. 

\begin{figure}
    \centering    \includegraphics{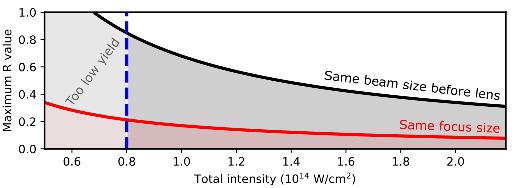}
    \caption{Range of two-color experiments possible (shaded area below lines), maximum achievable ratio for $R$ where both colors have the same beam diameter before the focusing lens (black) and for a configuration aiming for the same focus size of both colors (red). Above $\SI{0.8e14}{\watt\per\centi\meter\squared}$ (blue dashed line) the signal to noise ratio was sufficient. }
    \label{fig:results_range}
\end{figure}

$R=0$ and $R=1$ refer to only $\SI{1030}{\nano\meter}$ and $\SI{515}{\nano\meter}$, respectively, corresponding to single-color HHG in either case.
The total intensity $I_{\text{tot}}$ defines the values  of $R$ that can be taken.
$I_{\text{tot}}$ depends on the focusing geometries, pulse durations and pulse energies. In our experiment, the pulse duration for both colors is fixed and the pulse energies can be controlled by the attenuators. Since the beam sizes before the focusing lens are comparable, the focus size of the second harmonic is a factor of 2 smaller than for the fundamental [Eq.~(\ref{equ:w0})].
With the current configuration of parameters presented in Sec.~\ref{subsec:setup}, the possible values for $R$ are shown as the shaded grey area under the black line in Fig.~\ref{fig:results_range}. If another telescope would be inserted in, e.g., the $\SI{515}{\nano\meter}$ arm with the aim of having a comparable beamsize in the focus instead, the maximum value for $R$ would decrease rapidly (red).
For $I_{\text{tot}}<\SI{0.8e14}{\watt\per\centi\meter\squared}$, the signal to noise ratio was not sufficient anymore to evaluate the relative yield of the high harmonics in our experiment (blue dashed line). To still be able to access a large range for $R$, having a smaller second harmonic focus compared to the fundamental was deemed necessary. To the best of our knowledge, the impact of the relative focal  sizes in two or more color HHG has not been studied yet.  

\subsection{Integrated measurement platform}
All motorized components (waveplates for the attenuators, linear translation stages for focusing lens and camera I)  are controlled via a home-built control and measurement platform in Python. The cameras imaging the gas nozzle and the MCP, as well as the SLMs can be accessed within the same environment. For each image of the XUV spectrum, a variety of information is logged together with it. This includes the measured two-color phase, its  standard deviation, the settings of the waveplates defining the intensity ratio, and the settings of the SLMs and focusing lenses. With the help of this measurement platform,  different scan options can be selected that are then being conducted and saved in a completely automatized way, for example: 
\begin{itemize}
    \item relative phase over a $2\pi$ range
    \item relative intensity ratio while keeping the same total intensity
    \item focii position along the propagation direction
    \item focii position relative to each other
\end{itemize}
Thanks to this control system measurement times can be drastically shortened, avoiding complications due to long-term drifts,  thus improving the overall quality and reproducibility of the experiments. 

\section{Results}
\begin{figure}
    \centering
    \includegraphics{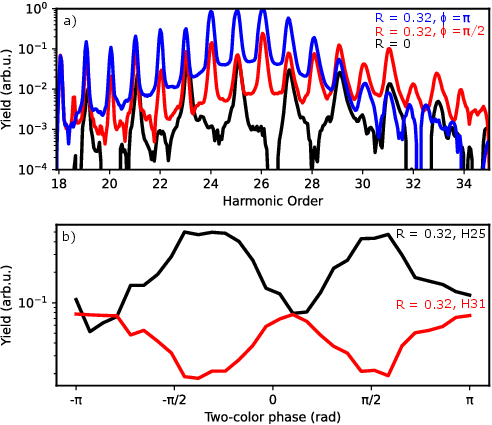}
    \caption{a) Results at $I_{\text{tot}}=\SI{0.8e14}{\watt\per\centi\meter\squared}$. Comparison of $R=0$ (black, $\SI{1030}{\nano\meter}$ only) and two-color at the same total intensity at different two-color phases for  $R=0.32$ (blue, red). b) Comparison of different harmonic yield oscillations over a $2\pi$ phase variation for harmonic 25 (black) and cut-off harmonic 31 (red).}
    \label{fig:results_harmonics}
\end{figure}
In Fig.~\ref{fig:results_harmonics}, results are shown showcasing the capabilities of our setup.
In this example, the total intensity of ${I_{\text{tot}}=\SI{0.8e14}{\watt\per\centi\meter\squared}}$ was kept constant while varying the intensity ratio between  $\SI{1030}{\nano\meter}$ and $\SI{515}{\nano\meter}$. In Fig.~\ref{fig:results_harmonics} a), the harmonic spectrum in argon at a backing pressure of $\SI{3}{\bar}$ is shown, for only the fundamental wavelength, ($R=0$, black). As expected, only odd harmonic orders are present. 
For the results shown in the red and blue curves, we keep the same total intensity but change the ratio to $R=0.32$, thus lowering the contribution of $\SI{1030}{\nano\meter}$ and adding $\SI{515}{\nano\meter}$. We consider two relative phases of $\phi=\pi$ (blue) and $\phi=\pi/2$ (red). We use the same parameters for $R$ and $\phi$ as in Fig.~\ref{fig:multicolor_schematics}, see Sec.~\ref{ssec:HHGintro}. Even harmonic orders appear in addition to the odd ones, with comparable intensities.
 By changing the phase, one can either move the cut-off to higher energies (red) or maximize  the lower and middle harmonic yields by more than one order of  magnitude (blue) compared to the single-color case (black). 
 
 In Fig.~\ref{fig:results_harmonics} b), the evolution of the yield of two different harmonics is shown for a full $2\pi$ phase scan. 
 Harmonic 25 (black) belongs to the HHG plateau, while harmonic  31 (red) is near the cut-off. Their oscillations are out of phase, reproducing previous studies \cite{shafir_atomic_2009, soifer_spatio-spectral_2013, mitra_suppression_2020, andiel_high-order_1999}. 
 
We noticed a continuous degradation of the $\SI{515}{\nano\meter}$ SLM after operating it on a daily basis over the course of one year, exposing it to a peak intensity of $\approx 10^9 \SI{}{\watt\per\centi\meter\squared}$ corresponding to an average intensity of $\SI{2}{\watt\per\centi\meter\squared}$. The slow deterioration resulted in a decrease of the reflection efficiency from $\SI{80}{\percent}$ initially to $\SI{70}{\percent}$ after one year. Further investigation of damage thresholds for SLMs using ultrashort laser pulses is therefore deemed necessary in the future.

\section{Conclusion and outlook}

In this work, we present a novel experimental setup designed for studying two-color HHG. A two-color interferometer is combined with SLMs for convenient spatial beam control but also for delay stabilization with a standard deviation of $<\SI{150}{\milli\radian}$ and a delay range corresponding to $2\pi$ of the second harmonic. Cameras for the beam profile and gas target allow the precise positioning of the two beams in respect to each other as well as the position of the geometrical focus in respect to the generation gas nozzle.

The strength of our setup lies in the extensive parameter control, allowing us to tune the relative intensity ratio $R$ and the phase difference between the two fields continuously, while keeping the total intensity fixed thanks to variable attenuators. 
This enables us to rigorously investigate properties, such as yield enhancement as a function of independent parameters and disentangle the contributions of different effects. 
Apart from the described two-color HHG, the setup additionally features unique possibilities for spatial beam shaping, e.g. top-hat beams and angular orbital momentum, with virtually unlimited possibilities to study details of single- and two-color HHG.

\begin{acknowledgments}
The authors acknowledge support from the Swedish Research Council (2013-8185, 2021-04691, 2022-03519, 2023-04603), the European Research Council (advanced grant QPAP, 884900), the Crafoord Foundation and the Knut and Alice Wallenberg Foundation. AL is partly supported by the Wallenberg Center for Quantum Technology (WACQT) funded by the Knut and Alice Wallenberg Foundation. 
\end{acknowledgments}

\section*{Data Availability Statement}
The data that support the findings of this study are available from the corresponding author upon reasonable request.


\bibliography{PAPER_article}

\end{document}